\documentclass[preprint,review,12pt]{elsarticle}

\usepackage{amssymb}
\usepackage{amsmath}

\usepackage{booktabs}
\usepackage{bbm}
\usepackage{multirow}
\usepackage{siunitx}
\usepackage{float}
\usepackage{tikz}
\usepackage{pgfplots}

\journal{Computer Speech \& Language}

\begin{document}

\begin{frontmatter}



\title{Toward Objective and Interpretable Prosody Evaluation in Text-to-Speech: A Linguistically Motivated Approach}


\author[1,2]{Cedric Chan\corref{cor1}} 
\ead{cedricc@upenn.edu}

\author[1]{Jianjing Kuang\corref{cor1}}
\ead{kuangj@upenn.edu}

\affiliation[1]{organization={Department of Linguistics, University of Pennsylvania},
            city={Philadelphia},
            state={PA},
            country={U.S.A.}}

\affiliation[2]{organization={Department of Computer and Information Science, University of Pennsylvania},
            city={Philadelphia},
            state={PA},
            country={U.S.A.}}

\cortext[cor1]{Corresponding author}

\begin{highlights}
\item Introduces a linguistically informed, objective, interpretable metrics for text-to-speech prosody evaluation
\item Employs a two-tier architecture linking discrete prosodic events with continuous realizations
\item Reveals model-specific prosodic weaknesses that traditional MOS tests cannot identify
\item Captures natural variability across speakers and acoustic cue dimensions
\end{highlights}

\begin{abstract}
Prosody is essential for speech technology, shaping comprehension, naturalness, and expressiveness. However, current text-to-speech (TTS) systems still struggle to accurately capture human-like prosodic variation, in part because existing evaluation methods for prosody remain limited. Traditional metrics like Mean Opinion Score (MOS) are resource-intensive, inconsistent, and offer little insight into why a system sounds unnatural. This study introduces a linguistically informed, semi-automatic framework for evaluating TTS prosody through a two-tier architecture that mirrors human prosodic organization. The method uses quantitative linguistic criteria to evaluate synthesized speech against human speech corpora across multiple acoustic dimensions. By integrating discrete and continuous prosodic measures, it provides objective and interpretable metrics of both event placement and cue realization, while accounting for the natural variability observed across speakers and prosodic cues. Results show strong correlations with perceptual MOS ratings while revealing model-specific weaknesses that traditional perceptual tests alone cannot capture. This approach provides a principled path toward diagnosing, benchmarking, and ultimately improving the prosodic naturalness of next-generation TTS systems.
\end{abstract}



\end{frontmatter}



\section{Introduction}

Transformative advancements in deep neural net (DNN) speech synthesis systems \cite{ren2019fastspeech, ren2020fastspeech} have produced TTS models that, particularly in short segments, are nearly indistinguishable from humans. While text inputs have continued to increase in length, content, and context, however, TTS models have yet to fully encapsulate the full range of human expression, inviting increased focus on modeling human prosody in speech technologies \cite{oh2024diffprosody, zhu2024metts, liu2024pe}.

More than just emotion, as it is frequently watered down into, prosody is essential for speech understanding. Prosody provides important cues for syntactic parsing, especially when structure is ambiguous \cite{millotte2007phrasal,wasow2005puzzle,snedeker2003using}. For example, in the sentence “Anna dressed the baby in the crib,” prosodic phrasing determines whether \textit{in the crib} describes \textit{where} the dressing took place or \textit{which} baby is being dressed. Such boundary placement cues allow speakers and listeners to correctly convey and recover the intended syntactic structure. Prosody also interacts with information structure to highlight what is new or important in discourse \cite{buring2013syntax, calhoun2010centrality}. For instance, emphasizing “\textit{JOHN} bought the book” signals that \textit{who} performed the action is the focus, while emphasizing “John bought the \textit{BOOK}” highlights \textit{what} was bought. The role of prosody spans several linguistic layers, from syntactic and semantic to even phonemic understandings of natural speech signals \cite{nespor2007prosodic, truckenbrodt1999relation, selkirk1984syntax, toivanen2002prosody}. Indeed, intonation carries pragmatic meaning beyond the literal words: what is said matters, but so does how it is said. Natural speech, therefore, should not only feature accurate pronunciation but also convey appropriate prosody for given contexts. 

Challengingly for TTS systems, prosody is complex. It is not solvable by, for example, adding SSML (Speech Synthesis Markup Language) \cite{taylor1997ssml} tags to the inputs of TTS models—something which, even if capable of encapsulating the full range of human prosody, adds a laborious step to what is intended to be an automated process. Rather, we know that prosody can be accidental or intentional, covert or overt, varying across a broad swath of linguistic and social contexts while maintaining the same emotional ``label'' in layman's terms \cite{wilson2006relevance}.

Problematically, the most common techniques for evaluating the prosodic accuracy of models remain resource-intensive and oftentimes opaque. These techniques can be broadly partitioned into subjective methods, which typically involve some sort of perception experiment where listeners rate or compare generated stimuli, and objective methods, which seek to score models purely based on the acoustic outputs.

\subsection{Subjective methods}

One of the most popular general measures of TTS quality—subjective or otherwise—is the Mean Opinion Score (MOS), which is calculated by conducting experiments that ask many participants to rate the ``naturalness'' of model outputs on a scale from 1 to 5, then taking the mean. Aside from being costly and time-intensive, MOS and similar evaluation methods have been found to be inconsistent across the literature and poorly defined, frequently leading to different results when performed under different circumstances \cite{chiang2023we}. For example, \cite{wester2015we} performed a meta-analysis of evaluations used at INTERSPEECH 2014 and found that more than 60\% of papers used fewer than 20 listeners for their evaluations. Meanwhile, they showed that at least 30 participants were necessary to enable a stable level of significance for MOS.

Another common technique is MUSHRA, in which participants evaluate several TTS models simultaneously along a sliding scale from 0 to 100 \cite{bs20151534}. The models are mixed in with an open reference produced by a human, as well as other lower-quality ``anchor'' references such as low-pass filtered speech samples. MUSHRA and similar techniques rely heavily, however, on a small sample of reference materials used in an artificial evaluation environment. Considering recent advances in TTS quality, this can prove problematic when synthetic outputs differ from the reference in a plausible manner. Low-quality anchors may also prove less useful in these settings.

To address some of the subjectivity and context-based limitations of direct rating tasks, paired comparison tests are also common for evaluating systems. For example, the classic AB preference test has participants choose between the tested system and a baseline using several different stimuli \cite{chevelu2015compare}. Based on these results, some sort of statistical model is then used to produce a final ranking or score. One example is the Bradley–Terry model (BTM) \cite{bradley1952rank}, which estimates latent competitiveness scores from pairwise comparisons. While providing a robust statistical foundation for comparing models, this class of evaluation metrics suffers a similar pitfall to MOS and MUSHRA, failing to provide useful linguistic details about the final score. A secondary limitation, which is in fact shared by all subjective methods, is that perception experiments can be resource-intensive, especially when performed at a scale that can ensure statistical significance. This is particularly true for paired comparison methods, where the number of pairings that must be tested increases quadratically in the number of evaluated models.

\subsection{Objective methods}

While subjective methods are still commonly recognized as the gold standard, there have been increasing attempts to address some of their limitations by using objective methods. For example, given the plethora of MOS rating data available, a natural idea would be to predict MOS scores directly from the TTS output using supervised machine learning models. Indeed, several attempts have been made in this general direction. MOSNet uses spectrograms as the input to predict MOS on a frame-by-frame basis, taking advantage of convolutional and recurrent layers in a bidirectional long short-term memory network to capture temporal and local information \cite{lo2019mosnet}. LDNet, which takes inspiration from MOSNet, additionally incorporates the listener's identity as an input, allowing for prediction for a specific listener \cite{huang2022ldnet}. Other models like SSL-MOS use pre-trained embeddings, rather than pure acoustic or spectral information as the input \cite{cooper2022generalization}. Some also include more specific linguistic features, such as $F_0$, POS tags, etc. \cite{vioni2023investigating}. Crucially, however, even if these models are able to successfully model MOS ratings, they necessarily possess the same limitations as the metric they mimic: inconsistency and linguistic opaqueness. 

Unsupervised models have also been used to calculate the probability of the naturalness of synthetic speech. \cite{falk2008towards}, for example, trains Hidden Markov models (HMMs) on natural speech, then calculates the log-likelihood of synthetic speech under those models. Performance, however, shows substantial gender differences (a separate HMM is used for each gender, using a simple $F_0$ frequency threshold to discriminate between them) and is largely limited to checking temporal features against scores like MOS on older, non-neural models. 

Word error rate (WER) is a third approach that looks at how well automatic speech recognition (ASR) systems recognize TTS outputs as a proxy for how well humans might, sometimes being correlated with human ratings \cite{seljan2013automatic}. Still, WER and similar techniques focus on segmental sequences and are generally insensitive to prosodic variation. Additionally, as TTS systems improve in quality, the focus in evaluation has shifted from the simpler task of understanding to the more complex one of employing accurate prosody.

A subset of objective methods evaluate speech using explicit linguistic cues. Rather than having a model predict a MOS score, these methods calculate scores directly from the linguistic features of the speech signal. One of the most common strategies to accomplish this is to compare synthetic speech data with a natural speech corpus. For example, \cite{kubichek1993mel} measures the Euclidean distance between the Mel-frequency cepstral coefficients of synthetic and reference speech, aligned via dynamic time warping. Similarly, \cite{suzuki2008automatic} calculates quantitative measures of rhythm and intonation based on acoustic features and compares them to a database of natural speech, combining the results into an overall score. 

Although these methods capture certain aspects of linguistic form, they are limited by their lack of sensitivity to variation. Frame-level acoustic comparisons assume that every utterance has a single optimal realization, unfairly penalizing valid prosodic variation that occur within the natural range of human expression. In practice, two speakers—or even two utterances by the same speaker—may express the same prosodic target with different pitch ranges, voice qualities, or timing patterns, all of which are perceptually valid. Evaluations based on rigid acoustic distances therefore risk rewarding uniformity rather than communicative adequacy.

\subsection{Current work: Linguistically informed objective metrics}

To address these limitations, our framework extends existing objective evaluation techniques in two key ways. First, we explicitly model human variation, incorporating the natural flexibility observed in prosodic realization to prevent over-penalization of legitimate variation in TTS outputs. In other words, our framework recognizes that there is not a single correct prosodic realization, but rather a range of acceptable patterns that convey the same communicative function. More importantly, we introduce evaluation using a two-tier framework that is grounded in prosodic theory.

An important insight from prosodic theory is that prosodic encodings of human natural speech are inherently two-layered: they consist of discrete structural targets and their continuous phonetic realizations. In intonational phonology (e.g., Autosegmental–Metrical theory), the discrete layer comprises categories such as pitch accents, phrase accents, and boundary tones, which define what events occur and where \cite{pierrehumbert1980phonology, ladd2008intonational}. These targets are then realized through continuous parameters—$F_0$ alignment, scaling, interpolation—modulated by duration, intensity, and voice quality adjustments. 

Computational models of human speech prosody,  though developed in different traditions, share the goal of linking abstract representations of prosodic events to their continuous acoustic realizations. The command–response model represents targets as underlying commands generating smooth contours \cite{fujisaki1984analysis}; the Tilt model parameterizes each event in shape and amplitude \cite{taylor2000analysis}; MOMEL/INTSINT stylizes contours by extracting sparse targets and interpolating between them \cite{hirst1993automatic}; the Target Approximation (qTA/PENTA) model further links target selection to communicative goals and models the dynamics of their realization \cite{prom2009modeling}.

What unites these approaches is the recognition that the continuous layer is not fixed: variation can come from multiple sources, including intrinsic speaker differences (e.g., pitch range, voice quality), contextual influences (e.g., syntax, discourse structure, and information status), and communicative intent (e.g., emphasis, affect). This means that evaluating prosody requires not only checking whether events are placed in the right locations, but also measuring how closely their continuous realization matches the range of natural variability observed in human speech. This dual perspective—structure plus execution—is the core principle guiding our two-tier evaluation.

Moreover, prosodic structure is realized through a rich, high-dimensional set of acoustic cues \cite{frazier2006prosodic}. This multidimensionality is evident in both phrasing and prominence, the two most important aspects of prosodic structure cross-linguistically. Phrasing is primarily related to durational cues such as word duration and pause duration, but also related to pitch cues (e.g., final lowering, reset, boundary tone) \cite{silverman1992standard}, and voice quality cues (e.g., creakiness) \cite{lee2015creaky,kuang2023boundary}. Prominence, meanwhile, can be variably realized through a number of acoustic cues: for example, in American English, stressed or accented words tend to have higher pitch, longer duration, tenser voice quality (stronger energy in the high-frequency region of the spectrum and greater periodicity), and greater intensity. And while traditional prosodic evaluations often only focus on pitch, intensity, and duration, recent studies have shown voice quality to contribute important acoustic-prosodic cues as well \cite{da2022effects, haderlein2011intelligibility}. Because speakers can use different combinations of cues to achieve the same communicative function, this cue multidimensionality thus contributes to an additional source of variation in prosodic realization. Our model therefore aims to capture a range of acoustic representations for prosodic structure, as well as their potential variability. Importantly, these cues are interpretable and linguistically meaningful. The goal of our approach is not only to provide objective evaluations for the prosodic naturalness of TTS systems, but also to diagnose and identify the limitations of state-of-the-art models, especially when the specific factors that make synthetic speech sound ``unnatural'' remain elusive to the untrained ear. 

Developing linguistically informed objective metrics for speech prosody is essential for advancing both scientific understanding and technological performance. Our proposed evaluation framework is designed to meet four key goals. First, it provides objective and reproducible measures of prosodic naturalness. Second, it adopts a two-tier architecture that reflects how human prosody operates—linking discrete structural events to their continuous phonetic realizations. Third, it accounts for variability across cue dimensions and individual speakers, capturing the natural diversity of prosodic expression. Finally, it is interpretable, enabling clear diagnosis of why and how synthetic speech diverges from human performance. To validate the robustness and perceptual relevance of these metrics, we compare model-based evaluations with human listener ratings, bridging quantitative analysis and perceptual judgment.

\section{Methods}

\subsection{Overview}

We extend and validate a new prosodic evaluation method we first introduced in \cite{chan2024exploring},  which automatically and objectively evaluates TTS outputs using acoustic measurements against a reference corpus of human speech. The method computes differences between TTS and human utterances of the same sentences across multiple acoustic measures, enabling interpretable, fine-grained analysis of model performance. Whereas perception MOS experiments incur significant costs whenever a model is updated, this method only has a one-time setup cost of collecting a test corpus, greatly decreasing evaluation costs and providing clarity into the specific acoustic domains in which particular models struggle. To capture the inherent variability of natural speech, evaluation is organized around two prosodic tiers—binary events (e.g., phrasing and prominence targets) and continuous signals (e.g., pitch and spectral trajectories).  Audio signals are force-aligned at the word level, and aggregate acoustic measurements are taken for each segment. These acoustic measurements serve as the basis for objective evaluation. Notably, rather than producing a single numerical score, the framework serves as an analytical tool for quantitatively identifying where and how TTS models diverge from human prosody. 


\subsubsection{Binary event evaluation}

As established in the linguistic literature based on natural human speech, the prosodic signal can be broken down into targets and interpolation. Here, ``prosodic events'' or simply ``events'' refer to those linguistic targets, including but not limited to pitch and phrase accents, boundary tones, and pauses. Indeed, the correct placement of such events—for example, accenting the right words and pausing in the appropriate places—is crucial in natural speech.  

The first tier therefore evaluates whether models place prosodic events on the same words where human speakers typically produce them. Events are automatically detected as local extrema in acoustic signals (e.g., peaks in the $F_0$ contour). We frame this as a binary classification task: for each word, the model predicts whether an event should occur. Because human productions vary, there is no single “correct” reference. To address this, we define two complementary criteria for correctness.

The simpler one looks solely at the proportion of speakers who agree with a model (i.e., place an event where the model does) at a particular point. We call this proportion, calculated for each word, the ``agreement score'' signal. More formally, we define a discrete signal comprising $n$ words as $x=(x_1, x_2, \dots, x_n), x_i\in\{0, 1\}$, where $x_i=1$ if there is an event at the $i$-th word, and $x_i=0$ otherwise. Our goal is to compare a machine signal $p=(p_1, p_2, \dots, p_n)$ against a set of human signals $\mathcal S=\{s_1,s_2,\dots,s_m\}$, where $s_i=(s_{i,1}, s_{i,2}, \dots, s_{i,n})$. Then, we define the agreement score of a model at the $i$-th word as
\begin{align}
    \alpha_i=\frac1m\sum_{j=1}^m\mathbbm1(p_i=s_{j,i})
\end{align}
where 
$$\mathbbm1(\phi)=\begin{cases}1 &\text{if $\phi$ is true}\\ 0 & \text{otherwise}\end{cases}$$
We can then establish a fixed threshold $c\in[0,1]$ that $\alpha_i$ must exceed for the model to be ``correct''—that is, the model is correct for word $i$ if and only if $\alpha_i\geq c$. In our experiments, we used $c=0.5$, a simple majority. Put simply, if the proportion of speakers who agree with the model (i.e., decide to put or not put an event where the model does) at a given point exceeds $c$, the model is considered correct for that event. 

We can now apply traditional binary classification evaluation metrics. To briefly illustrate, we define a slightly modified zero-one loss as follows:
\begin{align}
    \ell_{0/1}=\frac1n\sum_{i=1}^n\mathbbm1(\alpha_i<c)
    \label{equation:zero_one_loss}
\end{align}
where $n$ is the number of words. Similar definitions are used for precision, recall, and $F_1$:
   \begin{align*}
        {\rm Precision} &=\frac{\sum_{i=1}^n\mathbbm1(p_i=1\land \alpha_i\geq c)}{\sum_{i=1}^n\mathbbm1(p_i=1)} \\
        {\rm Recall} &=\frac{\sum_{i=1}^n\mathbbm1(p_i=1\land \alpha_i\geq c)}{\sum_{i=1}^n\mathbbm1\left[\frac1m\sum_{j=1}^m\mathbbm1(s_{j,i}=1)\geq c\right]}
    \end{align*}

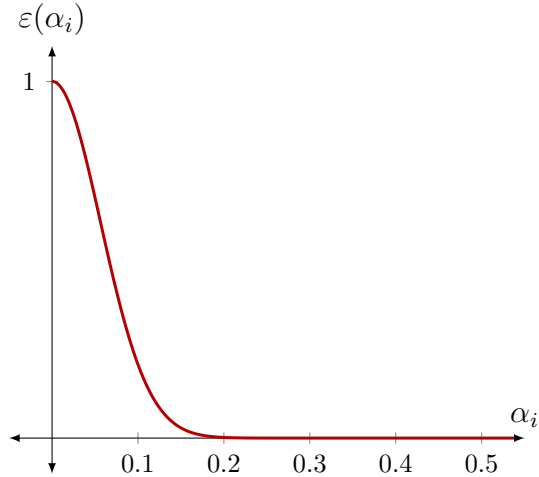
\begin{figure}[H]
        \centering
        \begin{tikzpicture}
            \begin{axis}[
                scale=1,
                xmax=0.5,ymax=1,
                xmin=0,
                axis lines=middle,
                restrict y to domain=-7:12,
                ytick={1},
                xlabel={$\alpha_i$},
                xlabel style={above},
                ylabel={$\varepsilon(\alpha_i)$},
                ylabel style={above},
                xticklabel={$\pgfmathprintnumber{\tick}$},
                yticklabel={$\pgfmathprintnumber{\tick}$},
                tick label style={font=\footnotesize},
                axis line style={latex-latex},
                enlargelimits,
            ]
                \addplot[domain=0:0.5375,samples=500,smooth,black!30!red,line width=0.4mm] {exp(-(4*pi*x)^2)};
            \end{axis}
        \end{tikzpicture}
        \caption{Plot of $\varepsilon$ as a function of $\alpha_i$.}
    \label{fig:decay}
    \end{figure}

It stands to reason, though, that being in the minority of human speakers does not preclude the possibility of a model being deemed natural. Indeed, the crux of the evaluation problem's difficulty is that natural variation permits multiple ``correct'' utterances. This leads us to the second method for determining ``correctness,'' which is a more lax version of the first. Rather than being strictly correct or incorrect, we assign a continuous correctness score $\varepsilon(\alpha_i)\in(0,1]$, where
\begin{align}
    \varepsilon(\alpha_i)=\exp\left[-(4\pi\alpha_i)^2\right]
\end{align}
As illustrated in Figure~\ref{fig:decay}, $\varepsilon$ is a generalized Gaussian function, with a smooth peak (i.e., high loss) when there's no or low agreement, and a rapid decline as agreement increases. Then, we can define a ``smooth'' loss $\ell_{0/1}^*$ by plugging in $\varepsilon(\alpha_i)$ for $\mathbbm1(\alpha_i<c)$ in Equation~\ref{equation:zero_one_loss}. Practically, this means that we're always awarding ``partial credit'' to models based on the proportion of speakers concurring with them: if all or almost all of the speakers agree with a model, it'll get all (or almost all) of the points, but as the percentage of people agreeing with it decays, its correctness will rapidly fall to zero. Thus, variability is enabled with a bias toward the majority.

\subsubsection{Event detection}

Thus far, we have skirted around any discussion of identifying events within signals, particularly imperfect, noisy signals that are collected in the field. Pause detection is straightforward: forced aligners automatically insert pause events (e.g., ``[SIL]'') between words. But particularly for events like peak $F_0$, fluctuations in the signal may lead to naive attempts to detect extraneous events.

We detect events automatically as peaks in the word-level signals of our selected acoustic features (consider, for example, how a pitch signal might exhibit a peak at an accented word). To reduce noise, we adopt the signal processing practice of using a moving-median filter threshold for peak selection \cite{kauppinen2002threshold}. Specifically, the median threshold $t_i$ for a signal $x$ is
\begin{gather*}
    t_i=\rho+\mathrm{median}\left(\{x_j\}_{j\in w}\right) \\
    w=\left[i-\left\lfloor\frac h2\right\rfloor, i+\left\lfloor\frac h2\right\rfloor\right]
\end{gather*}

\noindent where $h$ is the window size (in number of words) and $\rho$ is a fixed vertical shift to make the median threshold stricter. We used $h=7$ and $\rho=\frac12\cdot\mathrm{std}(x)$. Then, any peaks that are detected above this median threshold are selected as events. 

\subsubsection{Continuous signal evaluation}

While binary measures are our primary criterion for evaluating prosody, precise acoustic realizations are also crucial for simulating human speech. In the second class of evaluation metrics, we assess whether the ``continuous signals'' of TTS acoustic features fall within an appropriate range compared to natural human speech. This approach bears a closer resemblance to prior works, calculating the distance between acoustic measurements of human and synthetic utterances on some aligned basis. Here, we look at signals $x=(x_1, x_2, \dots, x_n), x_i\in\mathbb R$, where $x_i$ is an acoustic feature's measurement at the $i$-th word. Since each speaker uses the same script, all our word-aligned signals share the same dimensions, enabling evaluation via standard vector distance metrics. For example, we have a vector containing the $F_0$ measurement for each word in the human sample, and an analogous vector for the model, both with the same length. (As an aside, if we remove this assumption, a similar analysis is still possible, using alternate series comparison techniques like dynamic time warping.) 

Still, we run into the issue of comparing signals with a \textit{corpus} of human data; it's easy to compare a model with one human, but not all of them. To account for multiple speakers and the variation they exhibit, we look at ``normalized'' error: the mean of the squared $z$-scores of each element in the TTS signal. Formally, if we let $p=(p_1,\dots,p_n)$ be the TTS model's acoustic measurements and $\mathcal S_i=\{s_{j,i}\}_{j\in[1..m]}$ be the set of measurements of the $i$-th word among all human speakers (e.g., $\mathcal S_1$ is the set of measurements at the first word across all speakers), we can define error as follows:
\begin{align}
    \text{error}(p,\mathcal S)=\frac1n\sum_{i=1}^n\left(\frac{p_i-\overline{\mathcal S_i}}{\mathrm{std}(\mathcal S_i)}\right)^2
\end{align}
where $\overline{\mathcal S_i}$ is the mean of $S_i$ and $\mathrm{std}(\mathcal S_i)$ is its standard deviation. This is motivated by the fact that human speakers will likely have a high degree of agreement in certain places and low agreement elsewhere. By dividing out variation, our error metric weights model–human agreement based on whether people agree among themselves.

\subsubsection{Acoustic analysis}

In our experiments, human and TTS signals were force-aligned using Charsiu \cite{zhu2022charsiu}, a transformer-based aligner. Then, the following acoustic features were extracted on the word level using Praat: (1) duration (ms), including word duration and pause duration—important indicators for phrasing and temporal organization; (2) $F_0$ pitch (Hz), an important indicator for intonation, prominence, and phrasing); and (3) intensity (dB), which is important for prominence. In addition to these traditional prosodic features, we included three spectral measures that are important indicators for voice quality and prominence \cite{kuang2022effects, kuang2023boundary}: (4) alpha ratio, the energy difference between the 1–\SI{5}{\kilo\hertz} and 50–\SI{1}{\kilo\hertz} regions in the spectrum; (5) L1–L0, the difference between the $F_1$ (300–\SI{800}{\hertz}) and $F_0$ (0–\SI{300}{\hertz}) regions in the spectrum; and (5) cepstral peak prominence-smoothed (CPPS, dB). Before serving as the inputs of our evaluation metrics,  all measurements were $z$-score normalized by speaker and sentence. As an aside, we also performed a similar analysis at the phone, rather than word, level, but found inconsistent results due to high variation. Further, from the view of sentence-level prosody, the phone level emerged less relevant.

\subsection{Speech corpora}

We performed our analysis using a spoken corpus of Jane Austen’s \textit{Emma} (Volume II, Chapter 10), obtained through LibriVox \cite{librivox2005}. This dataset ensured a uniform narrative tone paired with moments of heightened emotion, especially in dialogue. Unlike spontaneous speech, these recordings contained relatively few disfluencies, a trait they share with synthetic speech. To facilitate meaningful comparisons with TTS outputs, our analysis centered on speakers 1 through 5, who exhibit North American English accents and characteristic female pitch ranges. The chapter comprises 136 sentences with an average length of 15.1 $\pm$ 11.4 words, ranging from 2 to 63.

Besides human speakers, we synthesized each sentence using the following five models, representing a range of capabilities across open-source and commercial models, and exhibiting the same accent characteristics as our corpus: Google TTS (\texttt{en-US-Studio-O}),\footnote{{https://cloud.google.com/text-to-speech?hl=en}} OpenAI TTS (\texttt{tts-1}; \texttt{nova}),\footnote{{https://platform.openai.com/docs/guides/text-to-speech}}  Amazon Polly (\texttt{Joanna}),\footnote{https://docs.aws.amazon.com/polly/} Microsoft Azure TTS (\texttt{en-US-Emma}),\footnote{{https://azure.microsoft.com/en-us/products/ai-services/ai-speech}} and VITS (\texttt{facebook/mms-tts-eng}) \cite{kim2021conditional}. For each of these models, we used the most recent generally available version as of February 2025.

\subsection{Human subject validation}

To demonstrate validity, we conducted traditional perception-based evaluation on the selected models with MOS and paired comparison tests. In both experiments, participants were instructed to ignore the meanings of sentences and focus only on naturalness. Our experiments were conducted online via PCIbex \cite{zehr2018penncontroller} surveys. Each experiment lasted approximately 30 minutes, although participants were able to replay audio samples throughout both experiments.

For the MOS experiment, participants were recruited through two pools: (1) 91 university students, and (2) 49 Prolific users (32 female, 17 male; mean age 40.84 $\pm$ 13.89). Prolific participants were required to have an 85–100\% approval rate, be based in the United States, and list English as their native language. Participants with outlier completion times and those who self-rated as being ``between focused and unfocused'' or worse were manually verified for attention. One participant was excluded because of irregular behavior. Participants were shown a random selection of 150 sentences across all speakers. As a control, roughly 15\% of stimuli presented to participants were human speech. For each sentence, participants rated the speaker's naturalness on the following scale \cite{shirali2023better}: (1) Completely unnatural, (2) Mostly unnatural, (3) In between unnatural and natural, (4) Mostly natural, or (5) Completely natural. Participants were additionally asked to answer ``Yes'' or ``No'' to the question, ``Do you believe this recording was spoken by a real person?''

In the pairwise comparison experiment, 97 participants were recruited from the university student population. Similar participation requirements were enforced. Each participant was presented with 115 random pairings of models speaking the same sentence, and was asked to choose which was more natural. Human speakers were not included in the stimuli.

Finally, to assess our framework’s treatment of natural variation, we also applied it to the human reference corpus itself using a leave-one-out design, evaluating each speaker against the remaining four. This allowed direct comparison between human and TTS prosodic performance under identical metrics.

\section{Results}

\subsection{Human Perceptual Evaluation}
\subsubsection{Overall human-likeness and Mean Opinion Scores}

Tables~\ref{tab:human_proportion} and \ref{tab:speaker_mos} summarize the results of the perceptual evaluation, presenting the proportion of listeners who judged each speaker as human and the MOS for overall naturalness, respectively. The MOS ratings followed a five-point scale \cite{shirali2023better}, where 1 corresponds to \textit{completely unnatural} and 5 to \textit{completely natural}. 

Notably, despite recent advances in TTS, both  metrics reveal that there remains a significant gap between synthetic and human performance in this reading task. For both MOS and the ratings on humanness, $t$-tests reveal that even the best-performing TTS model, OpenAI, was rated statistically lower than the worst-rated human speakers ($p=0.00011$). This underscores the persistent challenges that even the most advanced TTS systems face in achieving truly human-like naturalness. 

MOS scores in Table~\ref{tab:speaker_mos} suggest a general hierarchy of perceived quality (from best to worst): OpenAI, Google, Azure, VITS, and finally Polly. On the higher end, OpenAI and Google scored within the moderate naturalness range (3.0–4.0: borderline to mostly natural). VITS and Polly both occupied the lower end of the scale, in the range of ``mostly unnatural'' ($<3.0$), with VITS demonstrating a small but measurable advantage over Polly.  

\begin{table}[tbh]
  \caption{Proportion of participants who thought speaker was human, sorted.}
  \label{tab:human_proportion}
  \centering
  \begin{minipage}[b!]{0.45\linewidth}
    \centering
    \begin{tabular}{ l r }
      \toprule
      \textbf{Speaker} & \textbf{Proportion} \\
      \midrule
      S3           & 0.830 \\
      S1           & 0.812 \\
      S5           & 0.774 \\
      S4           & 0.770 \\
      S2           & 0.759 \\
      \bottomrule
    \end{tabular}
  \end{minipage} \hspace{4pt}
  \begin{minipage}[t]{0.45\linewidth}
    \centering
    \begin{tabular}{ l r }
      \toprule
      \textbf{Speaker} & \textbf{Proportion} \\
      \midrule
      OpenAI       & 0.625 \\
      Google       & 0.571 \\
      Azure        & 0.338 \\
      VITS         & 0.207 \\
      Polly        & 0.075 \\
      \bottomrule
    \end{tabular}
  \end{minipage}
\end{table}

\begin{table}[tbh]
  \caption{MOS of speakers, sorted.}
  \label{tab:speaker_mos}
  \centering
  \begin{minipage}[t]{0.45\linewidth}
    \centering
    \begin{tabular}{ l r }
      \toprule
      \textbf{Speaker} & \textbf{MOS} \\
      \midrule
      S3         & 4.14 $_{\pm\text{0.10}}$ \\
      S5         & 3.96 $_{\pm\text{0.11}}$ \\
      S2         & 3.91 $_{\pm\text{0.12}}$ \\
      S1         & 3.90 $_{\pm\text{0.11}}$ \\
      S4         & 3.77 $_{\pm\text{0.11}}$ \\
      \bottomrule
    \end{tabular}
  \end{minipage}
  \begin{minipage}[t]{0.45\linewidth}
    \centering
    \begin{tabular}{ l r }
      \toprule
      \textbf{Speaker} & \textbf{MOS} \\
      \midrule
      OpenAI     & 3.55 $_{\pm\text{0.05}}$ \\
      Google     & 3.41 $_{\pm\text{0.05}}$ \\
      Azure      & 2.80 $_{\pm\text{0.05}}$ \\
      VITS       & 2.09 $_{\pm\text{0.05}}$ \\
      Polly      & 1.84 $_{\pm\text{0.05}}$ \\
      \bottomrule
    \end{tabular}
  \end{minipage}
\end{table}

\subsubsection{Pairwise comparison}

\begin{figure}[t]
  \centering
  \includegraphics[width=0.55\linewidth]{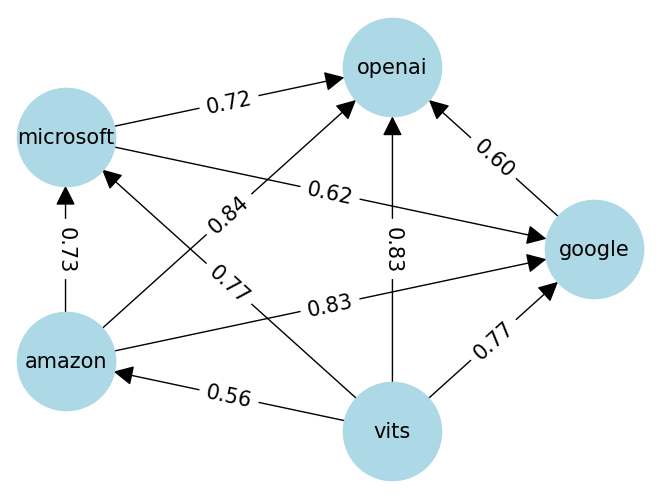}
  \caption{Proportions of the time models were deemed ``better'' than each other. Edges are directed toward the model that won the majority of the time.}
  \label{fig:win_props}
\end{figure}

While the MOS and human-likeness ratings summarize listeners’ general impressions, the results from the pairwise comparison task provide a more detailed view of relative preferences among TTS models. Figure~\ref{fig:win_props} models these pairwise comparisons as a directed graph, where each edge indicates the dominant model in a given pairing and edge weights correspond to the proportion of trials favoring that model. A clear transitive ranking emerges from the figure: OpenAI (no outgoing edges) occupies the highest rank, followed by Google (out-degree 1), Microsoft Azure (out-degree 2), Amazon Polly (out-degree 3), and VITS (out-degree 4). To validate this structure, we fitted a Bradley–Terry model (BTM) to the same dataset. As shown in Table~\ref{tab:btm_scores}, the BTM-derived scores reproduce the same ranking observed in Figure~\ref{fig:win_props}. 

Notably, the ranking from pairwise comparisons differs from that of MOS, with VITS rated lower than Polly. This suggests that when listeners compared samples directly, they preferred Polly slightly more often. The discrepancy highlights instability and context dependence of perceptual evaluations: even when rated by the same listeners, models were ranked differently when they were evaluated in isolation versus compared in pairs. This inconsistency underscores the limitations of traditional perceptual measures, calling into question the reliability of measures like MOS as standalone evaluation metrics.

Moreover, both MOS and pairwise comparisons provide only a coarse overall rating and cannot reveal the perceptual dimensions—such as timing precision, prosodic variability, or voice-quality cues—that drive listeners’ judgments. In other words, they only capture \textit{which} voices sound better, but not \textit{why}.  These findings motivate the need for more informative and interpretable evaluation approaches—such as the acoustic-prosodic analyses presented in the following section—that directly link measurable signal properties to perceptual outcomes.

\begin{table}[th]
  \caption{BTM scores of models, sorted}
  \label{tab:btm_scores}
  \setlength{\tabcolsep}{5pt}
  \centering
  \begin{tabular}{ l l l l l l }
    \toprule
    \textbf{} & \textbf{OpenAI} & \textbf{Google} & \textbf{Azure} & \textbf{Polly} & \textbf{VITS} \\
    \midrule
    \textbf{Score} & 0.946 & 0.575 & 0.136 & $-$0.780 & $-$0.877 \\
    \bottomrule
  \end{tabular}
\end{table}

\subsection{Objective prosodic evaluation}

\begin{figure}[t]
  \centering
  \includegraphics[width=0.65\linewidth]{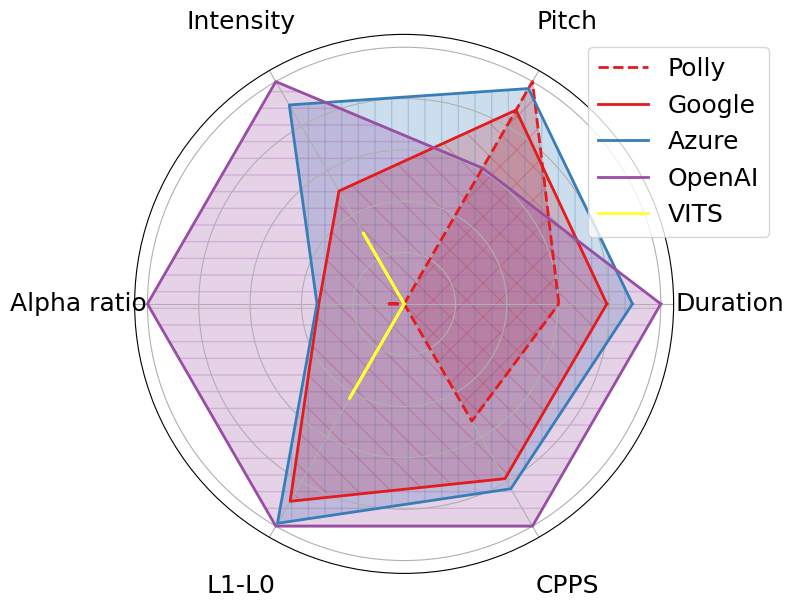}
  \caption{Performance based on $F_1$.}
  \label{fig:radar_f1}
\end{figure}

\begin{figure}[t]
  \centering
  \includegraphics[width=0.65\linewidth]{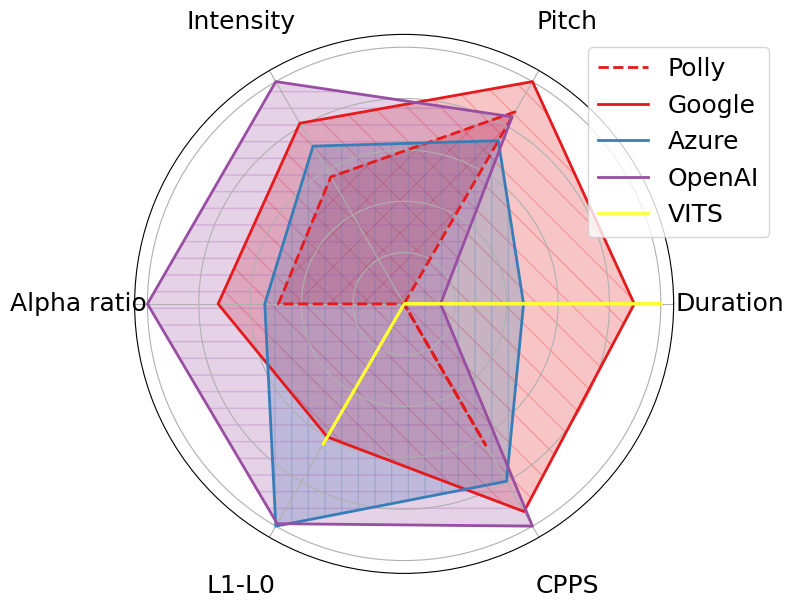}
  \caption{Performance based on normalized error. The complement of loss (i.e., $1-\mathrm{error}$) is used for easier visibility.}
  \label{fig:radar_z2}
\end{figure}

Figures~\ref{fig:radar_f1} and \ref{fig:radar_z2} summarize overall performance across all our measured acoustic-prosodic features using two different distance metrics. Full evaluation results for each measurement are in Appendix A. Figure \ref{fig:radar_f1} illustrates categorical (binary) accuracy using $F_1$ scores, which were min–max normalized for comparability, while Figure \ref{fig:radar_z2} depicts continuous performance using the complement of normalized error (1 – error), with higher values indicating closer approximations of human speech. Together, they capture complementary perspectives on prosodic control—discrete event placement versus continuous acoustic implementation — and reveal broadly consistent patterns across metrics.

Across both categorical and continuous metrics, OpenAI exhibits the most balanced and consistent prosodic control. It ranks highest in $F_1$ for duration (0.767), intensity (0.461), and spectral tilt (alpha ratio = 0.747), and maintains relatively low continuous errors across all cues, indicating accurate implementation of phrasing, prominence, and voice quality. However, its pitch accuracy ($F_1$ = 0.311; error = 0.582) slightly lags behind other cues, indicating persistent difficulty in reproducing natural intonation patterns and pitch accent placement. This limitation echoes the perceptual findings that, although OpenAI’s speech sounds more natural overall, its tonal modulation remains less human-like. 

Both Google Cloud and Microsoft Azure performed competitively across most prosodic dimensions.
While Azure achieved slightly higher scores on the binary metrics—reflecting more accurate placement of prosodic events such as prominence peaks or boundaries (e.g., duration $F_1$ = 0.723 vs. Google’s 0.684; spectral tilt $F_1$ = 0.642 vs. 0.641)—Google consistently showed lower normalized errors across the continuous metrics, indicating more precise acoustic realization of these cues once placed (e.g., duration error = 0.035 vs. Azure’s 0.042; pitch error = 0.468 vs. 0.660; spectral tilt error = 0.409 vs. 0.469). This distinction likely explains why Google’s speech was judged to sound more natural overall. Listeners are sensitive not only to whether prosodic events occur in the right locations but also to how smoothly and accurately those events are implemented acoustically. Fine-grained control over pitch movement, duration, and intensity variation contributes strongly to perceived naturalness.

Amazon Polly and VITS perform consistently below the top three systems across both categorical and continuous metrics. Polly shows its relative strength in pitch modeling, achieving higher categorical accuracy ($F_1$ = 0.418) and lower pitch error (0.566) than several other systems. Its performance on duration ($F_1$ = 0.611; error = 0.049) and CPPS ($F_1$ = 0.462; error = 0.620) is moderate, suggesting basic control over temporal structure and voice periodicity. However, Polly performs poorly on intensity ($F_1$ = 0.310; error = 0.576) and spectral tilt ($F_1$ = 0.599 for alpha ratio; 0.267 for L1–L0), suggesting limited variation in loudness and voice quality. VITS ranks the lowest among all systems, showing pervasive weaknesses in prosodic control across both categorical and continuous dimensions. It exhibits the lowest $F_1$ scores (e.g., duration = 0.375, pitch = 0.143, intensity = 0.358) and the largest continuous deviations (pitch error = 1.188; CPPS error = 0.757). The placement of phrasing and prominence is often inaccurate. Nonetheless, VITS shows some limited strengths. It achieves the lowest duration error among all models (0.033), suggesting consistent temporal pacing once a rhythm is established, even though its categorical accuracy for duration remains low ($F_1$ = 0.375).

Overall, the rankings derived from the acoustic–prosodic metrics using our proposed method are largely consistent with the perceptual results: OpenAI emerges as the most human-like system, followed by Google, Azure, Polly, and finally VITS.

\subsection{Human speaker prosodic self-validation}

\begin{table}[h!]
 \caption{$t$-tests for self-validation across all humans and models}
\centering
\begin{tabular}{l l r r r}
\toprule
\textbf{Feature} & \textbf{Metric} & \textbf{$t$-value} & \textbf{$p$-value} & \textbf{Winner} \\
\midrule
\multirow{3}{*}{Duration} 
    & $\ell_{0/1}^*$ & 8.521 & 1.13e-16 & Human \\
    & $F_1$          & $-$4.488 & 2.07e-05 & Human \\
    & Error          & 4.399 & 1.20e-05 & Human \\
\hline
\multirow{3}{*}{Pitch}
    & $\ell_{0/1}^*$ & 18.167 & 8.63e-60 & Human \\
    & $F_1$          & $-$7.420 & 3.37e-13 & Human \\
    & Error          & 13.278 & 5.68e-37 & Human \\
\hline
\multirow{3}{*}{Intensity}
    & $\ell_{0/1}^*$ & 17.275 & 3.59e-55 & Human \\
    & $F_1$          & $-$8.901 & 3.53e-18 & Human \\
    & Error          & 16.610 & 9.20e-55 & Human \\
\hline
\multirow{3}{*}{Alpha ratio}
    & $\ell_{0/1}^*$ & 14.544 & 1.29e-41 & Human \\
    & $F_1$          & $-$4.871 & 1.28e-06 & Human \\
    & Error          & 12.192 & 3.49e-32 & Human \\
\hline
\multirow{3}{*}{L1–L0}
    & $\ell_{0/1}^*$ & 18.013 & 5.50e-59 & Human \\
    & $F_1$          & $-$10.283 & 1.84e-23 & Human \\
    & Error          & 17.592 & 6.14e-61 & Human \\
\hline
\multirow{3}{*}{CPPS}
    & $\ell_{0/1}^*$ & 16.574 & 1.35e-51 & Human \\
    & $F_1$          & $-$8.897 & 2.82e-18 & Human \\
    & Error          & 16.329 & 1.36e-53 & Human \\
\bottomrule
\end{tabular}
\label{tab:merged_ttest}
\end{table}

To further validate the proposed acoustic–prosodic metrics, we conducted a self-validation analysis using human speech. If the metrics are meaningful, human speech should consistently outperform synthetic speech while still showing subtle variability among speakers.

Complete results for all speakers and acoustic measures are presented in Appendix B. Despite the decreased variation available for comparison with the self-validation task, human speakers score significantly better than the models do using our metrics. For duration, for example, the majority of speakers have an $F_1$ score of at least 0.9, whereas the best-performing model (OpenAI) scored merely 0.767. The most ``difficult'' feature to get right for both models and humans was L1–L0 spectral tilt, and even here the best model scores an $F_1$ of merely 0.375 compared to the worst human speaker's 0.521. The most striking measure is smoothed zero-one loss: across each of the five features, every human has a loss of 0.000, while no model ever achieves such a feat. As one of the metrics that intelligently encodes variation in the source dataset, it is a positive sign that our evaluation metrics can deftly handle multiple potentially correct utterances. 

More robustly, Table~\ref{tab:merged_ttest} shows the $t$-tests between all humans and all TTS models for each feature using smoothed zero-one loss, $F_1$, and normalized error. For each of the 18 shown settings, the human speakers perform statistically better than the TTS models do. Similar $t$-tests comparing the best TTS model against the worst human for each setting show the human performing better in 15 out of 18 settings for this more difficult task.

\section{Discussion}

Although state-of-the-art neural TTS systems now produce speech that is clear, smooth, and intelligible, our study demonstrates that a substantial gap remains between human and machine prosody. Both perceptual judgments and our acoustic–prosodic analyses converge on this conclusion: even the best-performing models fail to reproduce truly natural, human-like prosody. This gap calls for the need for effective and interpretable evaluation metrics—tools that can meaningfully quantify how far current systems remain from human prosodic behavior and, more importantly, why.

Traditional perceptual measures such as mean opinion scores (MOS) cannot serve this purpose. MOS offers a convenient overall impression of “naturalness,” but it collapses multiple perceptual dimensions—intonational appropriateness, rhythmic fluency, and voice quality—into a single number. As TTS intelligibility improves, these aggregated ratings become less diagnostic and more unstable, influenced by listener bias and task framing rather than systematic acoustic distinctions. As demonstrated in our experiments, the same participants produced inconsistent rankings across MOS and pairwise tasks, reflecting this contextual variability. Thus, perceptual testing is not sufficiently informative for identifying the specific prosodic mechanisms underlying perceived naturalness.

Building on linguistic insights regarding the dual-tier nature of prosody, our proposed evaluation framework fills this gap by combining both discrete and continuous metrics. The discrete tier captures the correct placement of prosodic events—phrasing, boundaries, and prominence—while the continuous tier measures how accurately those events are realized in the signal. This dual perspective enables a deeper diagnostic view: systems may succeed in event placement but fail in execution, or vice versa. For example, while Azure achieved higher binary accuracy than Google, it underperformed in continuous realization, leading to less natural-sounding speech. This finding illustrates that accurate event timing alone does not guarantee prosodic naturalness—proper fine-grained control over pitch movement, duration, and spectral balance is equally essential.

It is immediately clear that the overall trends observed in our perception experiments are reflected in the acoustic data. Our evaluation metrics uniquely highlight the varying strengths and weaknesses of TTS models across different acoustic dimensions. OpenAI’s model exhibited the most balanced overall performance, but its relatively weak pitch control suggests lingering challenges in generating natural intonation contours. Google Cloud, while achieving slightly lower categorical accuracy, demonstrated superior continuous precision, producing more stable cue realizations once prosodic events were correctly placed. Azure excelled in event placement yet lagged in voice-quality modulation and spectral variation, leading to overall lower naturalness. The lower-performing systems, Polly and VITS, showed greater inconsistencies in temporal alignment and pitch control, reflecting persistent difficulties in maintaining prosodic coherence within open-source architectures. The multidimensional results reveal that each system adopts a distinct prosodic strategy. Importantly, our metrics revealed that voice-quality measures—including alpha ratio, L1–L0, and CPPS—played a more significant role than previously assumed. Systems that exhibited richer spectral variability tended to sound more natural, even when their pitch trajectories were less accurate. These findings reinforce that prosodic expressiveness depends on the coordination of multiple acoustic dimensions—pitch, timing, and spectral quality—rather than on $F_0$ alone.

Moreover, a key advantage of this framework lies in its treatment of speaker variability. Prosody is inherently flexible: the same sentence may be realized with multiple acceptable contours, and speakers differ systematically in their use of pitch range, rhythm, and voice quality. Rigid reference-based metrics would penalize this natural variability as error. Our framework instead encodes variation within the human reference set, allowing models to be evaluated against the distribution of human utterances rather than a single canonical form. The human self-validation results confirm that the metrics behave sensibly: all human speakers scored near ceiling, with perfect smoothed loss ($\ell_{0/1}^*=0.000$) and $F_1$ scores exceeding 0.86 for duration, indicating that natural variability is preserved rather than punished. This ensures that the evaluation reflects genuine differences in prosodic control, not artifacts of inter-speaker diversity.

Finally, our results show that this flexibility extends beyond individual speakers to the multidimensional nature of prosodic realization. Among the prominence-related measures (intensity, pitch, spectral tilts, and CPPS), the high-performing models excelled in at least several of these dimensions. For example, while OpenAI did not score highly in pitch, it performed well in intensity, spectral tilt, and CPPS. This demonstrates that our evaluation metric objectively assesses synthesized speech naturalness while allowing for variability in the realization of prosodic prominence, as observed in natural speech. Different models, like human speakers, may use distinct acoustic strategies to achieve perceptually equivalent effects—an interpretive nuance that traditional single-dimensional metrics such as MOS cannot reveal. 

\section{Conclusions and future directions}

In this study, we introduced and validated a novel linguistically informed, semi-automatic prosodic evaluation technique for TTS models using a two-layered approach. This design reflects how human prosody operates—linking discrete categories such as phrasing and prominence to their continuous acoustic correlates—and allows both objective quantification and interpretive insight. 

The evaluation metrics used in this study align closely with the naturalness rankings produced by perception-based human experiments, but more significantly, provide not only a broad performance comparison but also insight into specific prosodic weaknesses of different TTS models. For example, some models needed more work on having appropriate voice quality throughout, while others needed work in accenting the correct words. The general framework for evaluation presented—separately evaluating continuous and binary signals against human speech corpora—importantly provides a better-defined path toward TTS improvement, not just relative to other models, but in line with objective linguistic features. 

Future work will extend the approach in several directions. Incorporating additional acoustic and spectral measures may capture finer-grained dimensions of prosodic expressiveness, while developing weighted composite indices could integrate multiple prosodic dimensions without sacrificing interpretability. Cross-linguistic applications and analyses of conversational or emotionally expressive speech will further test the framework’s generalizability. Finally, embedding these interpretable metrics directly into training or fine-tuning pipelines could transform them from diagnostic tools into guiding objectives for adaptive model optimization.

Together, these advances lay the groundwork for a transparent, linguistically principled evaluation standard that moves beyond mean opinion scores and toward a more systematic understanding of how expressive, human-like speech can be achieved in next-generation TTS models.

\appendix
\section{Full evaluation results for TTS models}

\begin{table}[H]
  \caption{Evaluation metrics for duration.}
  \label{tab:duration_metrics}
  \centering
  \setlength{\tabcolsep}{5pt} 
  \begin{tabular}{ l r r r r r r }
    \toprule
    \textbf{Speaker} & $\ell_{0/1}$ & $\ell_{0/1}^*$ & \textbf{Rec.} & \textbf{Prec.} & $F_1$ & \textbf{Err.} \\
    \midrule
        Polly    & 0.023 & 0.013 & 0.125 & 0.194 & 0.611 & 0.049 \\
    Google   & 0.017 & 0.007 & 0.375 & 0.417 & 0.684 & 0.035 \\
    Azure    & 0.013 & 0.007 & 0.196 & 0.773 & 0.723 & 0.042 \\
    OpenAI   & 0.014 & 0.006 & 0.657 & 0.540 & 0.767 & 0.047 \\
    VITS     & 0.021 & 0.013 & 0.140 & 0.227 & 0.375 & 0.033 \\

    \bottomrule
  \end{tabular}
\end{table}

\begin{table}[H]
  \caption{Evaluation metrics for mean pitch.}
  \label{tab:pitch_metrics}
  \centering
  \setlength{\tabcolsep}{5pt} 
  \begin{tabular}{ l r r r r r r }
    \toprule
    \textbf{Speaker} & $\ell_{0/1}$ & $\ell_{0/1}^*$ & \textbf{Rec.} & \textbf{Prec.} & $F_1$ & \textbf{Err.} \\
    \midrule
           Polly    & 0.124 & 0.038 & 0.514 & 0.257 & 0.418 & 0.566 \\
    Google   & 0.134 & 0.030 & 0.485 & 0.220 & 0.383 & 0.468 \\
    Azure    & 0.136 & 0.037 & 0.568 & 0.231 & 0.410 & 0.660 \\
    OpenAI   & 0.121 & 0.032 & 0.334 & 0.205 & 0.311 & 0.582 \\
    VITS     & 0.158 & 0.071 & 0.184 & 0.078 & 0.143 & 1.188 \\

    \bottomrule
  \end{tabular}
\end{table}

\begin{table}[H]
  \caption{Evaluation metrics on mean intensity}
  \label{tab:spl_metrics}
  \centering
  \setlength{\tabcolsep}{5pt} 
  \begin{tabular}{ l r r r r r r }
    \toprule
    \textbf{Speaker} & $\ell_{0/1}$ & $\ell_{0/1}^*$ & \textbf{Rec.} & \textbf{Prec.} & $F_1$ & \textbf{Err.} \\
    \midrule
          Polly    & 0.163 & 0.049 & 0.352 & 0.203 & 0.310 & 0.576 \\
    Google   & 0.154 & 0.041 & 0.479 & 0.251 & 0.387 & 0.498 \\
    Azure    & 0.143 & 0.034 & 0.507 & 0.304 & 0.446 & 0.532 \\
    OpenAI   & 0.126 & 0.023 & 0.494 & 0.342 & 0.461 & 0.438 \\
    VITS     & 0.180 & 0.063 & 0.440 & 0.229 & 0.358 & 0.760 \\

    \bottomrule
  \end{tabular}
\end{table}

\begin{table}[H]
  \caption{Evaluation metrics on spectral tilt (alpha ratio)}
  \label{tab:alpha_ratio_metrics}
  \centering
  \setlength{\tabcolsep}{5pt} 
  \begin{tabular}{ l r r r r r r }
    \toprule
    \textbf{Speaker} & $\ell_{0/1}$ & $\ell_{0/1}^*$ & \textbf{Rec.} & \textbf{Prec.} & $F_1$ & \textbf{Err.} \\
    \midrule
           Polly    & 0.132 & 0.042 & 0.675 & 0.507 & 0.599 & 0.487 \\
    Google   & 0.124 & 0.035 & 0.715 & 0.540 & 0.641 & 0.409 \\
    Azure    & 0.118 & 0.035 & 0.716 & 0.573 & 0.642 & 0.469 \\
    OpenAI   & 0.086 & 0.025 & 0.805 & 0.652 & 0.747 & 0.316 \\
    VITS     & 0.142 & 0.054 & 0.584 & 0.502 & 0.588 & 0.651 \\

    \bottomrule
  \end{tabular}
\end{table}

\begin{table}[H]
  \caption{Evaluation metrics on spectral tilt (L1–L0)}
  \label{tab:l1_l0_metrics}
  \centering
  \setlength{\tabcolsep}{5pt} 
  \begin{tabular}{ l r r r r r r }
    \toprule
    \textbf{Speaker} & $\ell_{0/1}$ & $\ell_{0/1}^*$ & \textbf{Rec.} & \textbf{Prec.} & $F_1$ & \textbf{Err.} \\
    \midrule
        Polly    & 0.213 & 0.067 & 0.372 & 0.166 & 0.267 & 0.873 \\
    Google   & 0.163 & 0.049 & 0.442 & 0.257 & 0.363 & 0.726 \\
    Azure    & 0.170 & 0.043 & 0.492 & 0.232 & 0.374 & 0.628 \\
    OpenAI   & 0.163 & 0.042 & 0.482 & 0.258 & 0.375 & 0.631 \\
    VITS     & 0.171 & 0.046 & 0.367 & 0.216 & 0.313 & 0.718 \\

    \bottomrule
  \end{tabular}
\end{table}

\begin{table}[H]
  \caption{Evaluation metrics for CPPS.}
  \label{tab:cpps_metrics}
  \centering
  \setlength{\tabcolsep}{5pt} 
  \begin{tabular}{ l r r r r r r }
    \toprule
    \textbf{Speaker} & $\ell_{0/1}$ & $\ell_{0/1}^*$ & \textbf{Rec.} & \textbf{Prec.} & $F_1$ & \textbf{Err.} \\
    \midrule
           Polly    & 0.170 & 0.044 & 0.503 & 0.373 & 0.462 & 0.620 \\
    Google   & 0.169 & 0.048 & 0.595 & 0.364 & 0.509 & 0.556 \\
    Azure    & 0.162 & 0.031 & 0.587 & 0.408 & 0.518 & 0.585 \\
    OpenAI   & 0.156 & 0.034 & 0.636 & 0.405 & 0.549 & 0.542 \\
    VITS     & 0.203 & 0.063 & 0.414 & 0.272 & 0.365 & 0.757 \\

    \bottomrule
  \end{tabular}
\end{table}

\section{Full evaluation results for human speakers}

\begin{table}[H]
  \caption{Human evaluation metrics for duration.}
  \label{tab:duration_metrics_human}
  \centering
  \setlength{\tabcolsep}{5pt}
  \begin{tabular}{ l r r r r r r }
    \toprule
    \textbf{Speaker} & $\ell_{0/1}$ & $\ell_{0/1}^*$ & \textbf{Rec.} & \textbf{Prec.} & $F_1$ & \textbf{Err.} \\
    \midrule
    S1 & 0.009 & 0.000 & 0.903 & 0.586 & 0.864 & 0.031 \\
    S2 & 0.007 & 0.000 & 0.815 & 0.677 & 0.907 & 0.024 \\
    S3 & 0.012 & 0.000 & 0.695 & 0.657 & 0.775 & 0.026 \\
    S4 & 0.008 & 0.000 & 0.750 & 0.721 & 0.905 & 0.025 \\
    S5 & 0.008 & 0.000 & 0.968 & 0.646 & 0.910 & 0.032 \\
    \bottomrule
  \end{tabular}
\end{table}

\begin{table}[H]
  \caption{Human evaluation metrics for mean pitch.}
  \label{tab:pitch_metrics_human}
  \centering
  \setlength{\tabcolsep}{5pt}
  \begin{tabular}{ l r r r r r r }
    \toprule
    \textbf{Speaker} & $\ell_{0/1}$ & $\ell_{0/1}^*$ & \textbf{Rec.} & \textbf{Prec.} & $F_1$ & \textbf{Err.} \\
    \midrule
    S1 & 0.108 & 0.000 & 0.729 & 0.310 & 0.543 & 0.384 \\
    S2 & 0.109 & 0.000 & 0.739 & 0.308 & 0.555 & 0.375 \\
    S3 & 0.097 & 0.000 & 0.666 & 0.315 & 0.535 & 0.382 \\
    S4 & 0.121 & 0.000 & 0.585 & 0.253 & 0.462 & 0.519 \\
    S5 & 0.090 & 0.000 & 0.769 & 0.359 & 0.588 & 0.353 \\
    \bottomrule
  \end{tabular}
\end{table}

\begin{table}[H]
  \caption{Human evaluation metrics for mean intensity.}
  \label{tab:spl_metrics_human}
  \centering
  \setlength{\tabcolsep}{5pt}
  \begin{tabular}{ l r r r r r r }
    \toprule
    \textbf{Speaker} & $\ell_{0/1}$ & $\ell_{0/1}^*$ & \textbf{Rec.} & \textbf{Prec.} & $F_1$ & \textbf{Err.} \\
    \midrule
    S1 & 0.103 & 0.000 & 0.705 & 0.423 & 0.597 & 0.339 \\
    S2 & 0.086 & 0.000 & 0.834 & 0.493 & 0.689 & 0.235 \\
    S3 & 0.125 & 0.000 & 0.643 & 0.350 & 0.512 & 0.326 \\
    S4 & 0.099 & 0.000 & 0.724 & 0.416 & 0.589 & 0.321 \\
    S5 & 0.100 & 0.000 & 0.692 & 0.447 & 0.606 & 0.294 \\
    \bottomrule
  \end{tabular}
\end{table}

\begin{table}[H]
  \caption{Human evaluation metrics on spectral tilt (alpha ratio).}
  \label{tab:alpha_ratio_metrics_human}
  \centering
  \setlength{\tabcolsep}{5pt}
  \begin{tabular}{ l r r r r r r }
    \toprule
    \textbf{Speaker} & $\ell_{0/1}$ & $\ell_{0/1}^*$ & \textbf{Rec.} & \textbf{Prec.} & $F_1$ & \textbf{Err.} \\
    \midrule
    S1 & 0.120 & 0.000 & 0.644 & 0.593 & 0.633 & 0.369 \\
    S2 & 0.079 & 0.000 & 0.828 & 0.653 & 0.749 & 0.296 \\
    S3 & 0.098 & 0.000 & 0.741 & 0.608 & 0.703 & 0.310 \\
    S4 & 0.074 & 0.000 & 0.854 & 0.675 & 0.762 & 0.287 \\
    S5 & 0.059 & 0.000 & 0.888 & 0.720 & 0.806 & 0.207 \\
    \bottomrule
  \end{tabular}
\end{table}

\begin{table}[H]
  \caption{Human evaluation metrics on spectral tilt (L1--L0).}
  \label{tab:l1_l0_metrics_human}
  \centering
  \setlength{\tabcolsep}{5pt}
  \begin{tabular}{ l r r r r r r }
    \toprule
    \textbf{Speaker} & $\ell_{0/1}$ & $\ell_{0/1}^*$ & \textbf{Rec.} & \textbf{Prec.} & $F_1$ & \textbf{Err.} \\
    \midrule
    S1 & 0.102 & 0.000 & 0.618 & 0.409 & 0.521 & 0.422 \\
    S2 & 0.123 & 0.000 & 0.771 & 0.392 & 0.593 & 0.384 \\
    S3 & 0.119 & 0.000 & 0.648 & 0.361 & 0.527 & 0.458 \\
    S4 & 0.124 & 0.000 & 0.699 & 0.363 & 0.525 & 0.518 \\
    S5 & 0.109 & 0.000 & 0.838 & 0.441 & 0.641 & 0.413 \\
    \bottomrule
  \end{tabular}
\end{table}

\begin{table}[H]
  \caption{Human evaluation metrics for CPPS.}
  \label{tab:cpps_metrics_human}
  \centering
  \setlength{\tabcolsep}{5pt}
  \begin{tabular}{ l r r r r r r }
    \toprule
    \textbf{Speaker} & $\ell_{0/1}$ & $\ell_{0/1}^*$ & \textbf{Rec.} & \textbf{Prec.} & $F_1$ & \textbf{Err.} \\
    \midrule
    S1 & 0.103 & 0.000 & 0.724 & 0.510 & 0.646 & 0.359 \\
    S2 & 0.093 & 0.000 & 0.801 & 0.553 & 0.715 & 0.304 \\
    S3 & 0.104 & 0.000 & 0.772 & 0.554 & 0.679 & 0.349 \\
    S4 & 0.136 & 0.000 & 0.677 & 0.440 & 0.587 & 0.447 \\
    S5 & 0.119 & 0.000 & 0.724 & 0.516 & 0.661 & 0.360 \\
    \bottomrule
  \end{tabular}
\end{table}

\begin{table}[H]
 \caption{$t$-tests for self-validation for best model vs. worst human}
\centering
\begin{tabular}{l l l r r r r}
\toprule
\textbf{Feature} & \textbf{Metric} & \textbf{TTS} & \textbf{Human} & \textbf{$t$-value} & \textbf{$p$-value} & \textbf{Winner} \\
\midrule
\multirow{3}{*}{Duration}
    & $\ell_{0/1}^*$ & OpenAI & S1 & 3.076 & 2.56e-03 & Human \\
    & $F_1$          & OpenAI & S3 & $-$0.104 & 9.17e-01 & Human \\
    & Error          & VITS   & S5 & 0.147 & 8.83e-01 & Human \\
\hline
\multirow{3}{*}{Pitch}
    & $\ell_{0/1}^*$ & Google & S4 & 6.977 & 1.56e-10 & Human \\
    & $F_1$          & Polly  & S4 & $-$0.712 & 4.77e-01 & Human \\
    & Error          & Google & S4 & $-$1.527 & 1.28e-01 & TTS \\
\hline
\multirow{3}{*}{Intensity}
    & $\ell_{0/1}^*$ & OpenAI & S3 & 5.750 & 6.14e-08 & Human \\
    & $F_1$          & OpenAI & S3 & $-$0.943 & 3.47e-01 & Human \\
    & Error          & OpenAI & S1 & 3.388 & 8.31e-04 & Human \\
\hline
\multirow{3}{*}{Alpha ratio}
    & $\ell_{0/1}^*$ & OpenAI & S1 & 5.626 & 1.09e-07 & Human \\
    & $F_1$          & OpenAI & S1 & 2.791 & 5.78e-03 & TTS \\
    & Error          & OpenAI & S1 & $-$1.906 & 5.79e-02 & TTS \\
\hline
\multirow{3}{*}{L1–L0}
    & $\ell_{0/1}^*$ & OpenAI & S4 & 7.181 & 4.90e-11 & Human \\
    & $F_1$          & OpenAI & S1 & $-$2.806 & 5.62e-03 & Human \\
    & Error          & Azure  & S4 & 3.437 & 6.88e-04 & Human \\
\hline
\multirow{3}{*}{CPPS}
    & $\ell_{0/1}^*$ & Azure  & S4 & 6.937 & 1.78e-10 & Human \\
    & $F_1$          & OpenAI & S4 & $-$0.876 & 3.82e-01 & Human \\
    & Error          & OpenAI & S4 & 2.892 & 4.21e-03 & Human \\
\bottomrule
\end{tabular}
\label{tab:best_worst_ttest}
\end{table}

\bibliographystyle{elsarticle-num} 
\bibliography{mybib}






\end{document}